\documentclass[aps,prl,twocolumn,preprintnumbers]{revtex4-1}
\usepackage{amsmath,amssymb,epsfig}
\usepackage{graphicx,floatflt,subfigure}
\usepackage{times}
\usepackage{latexsym}
\usepackage{multirow}
\usepackage{url}
\usepackage{color}

\def\beq{\begin{equation}}
\def\eeq{\end{equation}}
\def\bey{\begin{eqnarray}}
\def\eey{\end{eqnarray}}
\def\Ptmiss{\not{\hbox{\kern-3pt $E_T$}}}

\newcommand{\mathsym}[1]{{}}

\newcommand{\tmop}[1]{\ensuremath{\operatorname{#1}}}

\DeclareMathOperator{\gev}{GeV}

\begin{document}

\thispagestyle{empty}

\title{ Top Channel for Early SUSY Discovery at the LHC}
\author{Gordon L. Kane$^1$, Eric Kuflik$^1$,  Ran Lu$^1$,  Lian-Tao Wang$^{2,3}$}
\address{\textsuperscript{1} Michigan Center for Theoretical Physics, University of Michigan, Ann Arbor, MI 48109\\
\textsuperscript{2}  Department of Physics, Princeton University, Princeton, NJ 08540, USA \\
\textsuperscript{3} Department of Physics and Enrico Fermi Institute, University of Chicago, Chicago, IL 60637, USA
}

\date{\today}

\begin{abstract}
In recent years many models of supersymmetry have implied a large production rate for events  including a high multiplicity of third generation quarks, such as four top quarks. It is arguably the best-motivated channel for early LHC discovery.
A particular example  is  generic string theories compactified to four dimensions with stabilized moduli which typically have multi-TeV squarks and lighter gluinos (below a TeV) with a large pair production rate and large branching ratios to four tops. We update and sharpen the analysis 4-top signals and background to 7 TeV LHC energy. For 1 fb$^{-1}$ integrated luminosity, gluinos up to about 650 GeV in mass can be detected, with larger masses accessible for higher luminosities or at higher energies. More than one signature is likely to be accessible, with one charged lepton plus two or more b-jets, and/or same-sign dileptons plus b-jets being the best channels. A non-Standard Model signal from counting is robust, and provides information on the gluino mass, cross section, and spin. \end{abstract}

\preprint{MCTP-11-00}
\maketitle

\section{Introduction}

The Large Hadron Collider (LHC) is likely to accumulate significant amounts of data in 2011. While the detector groups will be sensitive to many ways new physics could appear, it is not possible to focus equally on all possible interesting signatures, so it is valuable to examine well-motivated  channels that may yield results at the initial LHC energies and luminosities. In recent years it has increasingly been recognized that considerations of new physics point toward top-quark and bottom-quark rich final states at the Large Hadron Collider (LHC), as naturalness of electroweak symmetry breaking (EWSB) typically requires the existence of a top partner to cancel the quadratic divergences in the Standard Model (SM).

Supersymmetry implies the existence of a top partner that cancels quadratic divergences. Supersymmetry also introduces a partner for the gluon, the gluino, in the low energy spectrum. At proton colliders, pair production of gluinos, and consequently their decay products, typically become the main channel of supersymmetric signals. Models with light top partners, are common and they imply that a typical signature of production of the gluino will be multiple top quarks in the final states \cite{lightgluino}.

In the rest of this paper, we will study this signature of low energy supersymmetry with light gluinos, focusing on the well-motivated scenario in which the squarks are considerably heavier than the gluino and the third generation squarks are lighter than those of the first two generations. In this case, the gluino will dominantly decay into top and/or bottom quarks. Earlier some of us, along with Acharya, Grajek, and Suruliz \cite{14TeVPaper} studied such processes in detail for the 14 TeV LHC. In this paper we update the study for early LHC at 7 TeV, and focus on the significant reach and robustness of a signal with the number of events from 1 fb$^{-1}.$

This scenario is a generic possibility from the point of view of many SUSY models \cite{models}. 
Heavier squark masses are often preferred due to constraints from flavor changing neutral currents and CP violation. Even more generally, when embedding low energy supersymmetry into a string theory, moduli stabilization and cosmological constraints imply that moduli masses and gravitno mass, and consequently scalar masses \cite{moduli}, must be larger than about 20 TeV \cite{modulibound}. Then, standard renormalization group (RG) running of scalar masses from the unification scale down to the electroweak scale will push the third generation squark masses significantly lower than those of the other generations.  In most cases this turns out to be right handed stop squark.  Alternative models leading to multi-top final states, and corresponding anaylsis approaches,  have been studied \cite{OtherStudies} (See Ref.~\cite{14TeVPaper} for a more extensive list.).

The gluino decays via virtual squarks to $q\bar{q}\chi_1^0$ or $q\bar{q}\chi_1^{\pm}$. Since the rate for a given diagram scales as the virtual squark mass to the $-4$ power from the propagator, the lightest squarks dominate. Therefore, we are led to consider decay channels $\tilde{g}\rightarrow t\bar{t}\tilde{N}$, $\tilde{g}\rightarrow t\bar{b}\tilde{C}^{-}$, and $\tilde{g} \rightarrow b\bar{b}\tilde{N}$. Decays of multiple top quarks lead to b-rich and lepton rich final states, and give excellent potential for early discovery. In fact, we show that significant excesses can be observed at the early LHC-7 TeV. For example, gluino masses larger than 600 GeV can be discovered in the single-lepton plus 4 b-jets channel.

We carry out our study on several benchmark models. To study the reach of gluino pair production, with decays into third generation squarks, a detailed scan of the parameter space involving the gluino mass and LSP mass, for different branching ratios, is performed. We emphasize that the goal of this study is to demonstrate that gluino pair production with decays via third generation squarks provides an ideal channel for early discovery at the LHC, since it leads to lepton and $b$-quark rich final states.

\section{Benchmark Models}
Three benchmark models are considered which will form the basis for the numerical scan discussed below. The model parameters and relevant decay branching ratios are shown in Table~\ref{tab:model}. Model A is a simple example of multi-top physics. The spectrum would have a stop much lighter than the other squarks, and therefore gluino pair production always produces four tops in the final state. Model B is designed to include the decay channel $\tilde{g} \rightarrow b \bar{b} \chi_1^0$, which will result if the sbottom is also lighter than the first two generation squarks, and $m_{\tilde{t}} \sim m_{\tilde{b}}$. Model B is observably different than Model A, while somewhat more difficult to discover. These models have a Bino-like LSP. In Model C, the Wino is the LSP, and is approximately degenerate with the lightest chargino, which is also Wino-like. It is designed to further include a chargino in the decay chain, which allows the decay $\tilde{g}\rightarrow t\overline{b} \chi_1^+$. Since the charged Wino is approximately degenerate with the wino LSP, it appears only as missing energy; though if one focuses on the signal events the chargino stub \cite{Feng:1999fu} can probably be seen in the vertex detector.
\begin{table}[t!]
\begin{center}
\begin{tabular}{|c|c|c|c|}
\hline
 &  \multicolumn{3}{c|}{Branching ratios}  \\
\hline
 &  $\tilde{g}\rightarrow t\bar{t} \chi_1^0 $ & $\tilde{g}\rightarrow
b\bar{b} \chi_1^0 $ & $\tilde{g}\rightarrow t \bar{b} \chi_1^+ +h.c.$  \\
\hline
A & 1 & 0 & 0  \\
\hline
B & 0.5 & 0.5 & 0 \\
\hline
C& .08 & 0.22 & 0.7  \\
\hline
\end{tabular}
\caption{\label{tab:model} Relevant branching ratios for the benchmark models considered in this paper. The models A and B have bino LSP. In Model C, the lightest neutralino and lightest chargino are both winos.  In all models the first two generation squark masses are taken to be 8 TeV. The third generation is taken to be somewhat lighter and is chosen to generate the required branching ratios of the model. }
\label{tab:model}
\end{center}
\end{table}

The three models are taken as a basis for 3 seperate numerical scans, where $m_{\tilde{g}}$ and $m_{LSP}$, are varied while the branching ratios are fixed, as shown in Table~\ref{tab:model}. In particular, scans in model A  and model B varied $m_{\tilde{g}}$ and $m_{\text{LSP} }=m_{\chi _{1}^{0}}$. while scan in model C
% wit fixing $Br(\tilde{g}\rightarrow t\overline{t}\chi _{1}^{0})=1$, Scan B varied $m_{\tilde{g}}$ and $m_{\text{LSP}}=m_{\chi _{1}^{0}}$ while fixing $Br(\tilde{g}\rightarrow t\overline{t}\chi _{1}^{0})=Br(\tilde{g}\rightarrow b\overline{b}\chi _{1}^{0})=0.5$, and Scan C
  varied $m_{\tilde{g}}$ and $m_{\text{LSP}}=m_{\chi _{1}^{0}}\simeq m_{\chi _{1}^{\pm }}$.
%   while fixing $Br(\tilde{g}\rightarrow t \bar{b} \chi_1^+ h.c)=0.7$, $Br(\tilde{g}\rightarrow t\overline{t}\chi _{1}^{0})=0.08$, and $Br(\tilde{g}\rightarrow b\overline{b}\chi _{1}^{0})=0.22$.

\section{Signal Isolation and Backgrounds}

The relatively large $b$-jet and lepton multiplicity associated with multiple top production provide for potentially striking signatures that are easily distinguishable above the expected SM background. By requesting multiple $b$-tagged jets and at least one lepton, it is possible to achieve signal significance $S/\sqrt{B}>5$ for 1 fb$^{-1}$ of integrated luminosity.

The most significant backgrounds from the SM for final states with many $b$-jets, several isolated leptons and missing energy, are from top pair production, $ t\bar{t}$. The expected cross-section at the LHC for 7-TeV center-of-mass energy is $\sigma = 164$pb (NLO) \cite{SMttAtlas}. Also included in the analysis are a set of SM backgrounds involving associated production of gauge bosons with third generation quarks. These contribute less significantly to the backgrounds than $t\bar{t}$, but can contribute to signals with high lepton multiplicity.  All background sources considered, and their respective cross sections are given in Table~\ref{tab:SMBCrossSections}. With the exception of the $t\bar{t }$ cross section, we increased all SM background cross sections by a factor of 2, to account for possible K-factor from NLO corrections. Since the relevant backgrounds for the channels considered end up small (Table \ref{tab:SMBCrossSections}), uncertainties in the cross section are not important.

All background event samples were produced with Madgraph v.4 \cite{madgraph}, while the parton shower and hadronization were done by Pythia 6.4 \cite{pythia}. Additional hard jets (up to three) were generated via Madgraph, while the MLM \cite{mlm} matching scheme implemented in Madgraph was used to match these jets to the ones produced in the Pythia showers. The events were then passed through the PGS-4 \cite{PGS4} detector simulators with parameters chosen to mimic a generic ATLAS type detector. The b-tagging efficiency was changed to more closely match the expected efficiencies at ATLAS \cite{btagatlas}. For $b$-jets with $50 \mbox{ GeV} \lesssim p_T\lesssim 200 \mbox{ GeV}$, which is typical of the $b$-jets in the signal, the efficiency is approximately 60\% for tagging a $b$-quark.

\begin{table}[t!]
\label{SignalBackground}
\begin{tabular}{|c|c|c|c|c|}
  \hline 
  Process & $\sigma$ [fb] & $\sigma_{L 1} [\tmop{fb}]$ & $\sigma_1 $[fb] & $\sigma_2 $[fb]\\
  \hline
  $b \bar{b} + \gamma / Z + \tmop{jets}$ & $4.69 \times 10^5$ & $1.41 \times
  10^4$ & 34.0 & 107.8\\
  \hline
  $b \bar{b} + W^{\pm} + \tmop{jets}$ & $2.41 \times 10^4$ & $5.39 \times
  10^2$ & 7.71 & 13.3\\
  \hline
  $t \bar{t} + \gamma / Z + \tmop{jets}$ & $1.54 \times 10^3$ & $7.69 \times
  10^2$ & 42.3 & 95.4\\
  \hline
  $t \bar{t} + W^{\pm} + \tmop{jets}$ & $2.25 \times 10^2$ & $1.31 \times
  10^2$ & 14.3 & 27.6\\
  \hline
  $t \bar{b} + \gamma / Z + \tmop{jets} + h.c.$ & $1.34 \times 10^3$ & $8.09
  \times 10^2$ & 7.37 & 26.6\\
  \hline
  $b \bar{b} + V V + \tmop{jets}$ & $1.14 \times 10^3$ & $2.33 \times 10^2$ &
  1.45 & 3.94\\
  \hline
  $t \bar{t} + \tmop{jets}$ & $1.60 \times 10^5$ & $6.60 \times 10^4$ &
  2076.7 & 5905.6\\
  \hline
  $V V + \tmop{jets}$ & $1.03 \times 10^5$ & $1.03 \times 10^5$ & 108.6 & 377.7\\
  \hline
  \hline
  Model A & $1.19 \times 10^3$ & $9.48 \times 10^2$ & 403.8 & 508.1\\
  \hline
  Model B & $1.19 \times 10^3$ & $1.03 \times 10^3$ & 505.2 & 703.1\\
  \hline
  Model C & $1.19 \times 10^3$ & $5.80 \times 10^2$ & 300.5 & 420.5\\
  \hline
\end{tabular}
\caption{Cross sections for production of signal and backgrounds. The first
column gives the total production cross section. The second gives the cross
section after the L1 triggers defined in PGS-4 (see text).  The remaining columns
give the cross section after selection cuts in Eq.~\ref{cut1} and Eq.~\ref{cut2}, with an additional missing energy (MET) requirement, ${\not}E_{T} \ge100$ GeV. The $b\bar{b}+\tmop{jets}$ and $b\bar{b}b\bar{b}$-inclusive backgrounds  have been considered, and after the applying the selection cuts in Eqs.~\ref{cut1}-\ref{cut2} and requiring at least one lepton, the number of events are negligible in the $\{b,\ell \}$ channels considered here.  
In this table, we set $m_{\tilde{g}} = 500$ GeV and $m_{LSP} = 100 $ GeV.}
\label{tab:SMBCrossSections}
\end{table}

The signal event samples, for gluino pair production and decay, were produced using Pythia 6.4 and have been passed through the same PGS-4 detector simulation. Basic muon isolation was applied to all samples. To reduce the number of backgrounds events are required to pass the L1-triggers as defined by PGS. 
We also display the effect of two possible additional  selection cuts, together with the additional requirement ${\not}E_{T} \ge100$ GeV,
\begin{eqnarray}
\text{\emph{cut-1}}: n_j(p_{T}\ge 50 \gev) \ge 4 \label{cut1}\\
\text{\emph{cut-2}}: n_j(p_{T}\ge 30 \gev) \ge 4 \label{cut2}
\end{eqnarray}
in the last two columns of Table~\ref{tab:SMBCrossSections}. 
The second cut (weaker than the first) is optimal for discovery signatures, such as the same-sign dilepton signature, that have relatively small SM backgrounds.

Next, the signal is searched for in multi $b$-jet ($n_b=2,3,4$) and multi lepton channels ($1\ell,SS,OS,3\ell$). All objects are required to have a minimum $p_T$ of 20 GeV.  Same sign (SS) and opposite sign (OS) di-leptons are separated as they can have different origins and sizes. We will use the possible excess in these channels to assess the discovery potential.  Table \ref{tab:counts} shows the expected number of events from the SM background as classified according to the number of $b$-tagged jets and isolated leptons in the event.

Table \ref{tab:counts}  shows the expected number of signal events with $b$-tagged jets and  isolated leptons  for the three benchmark models.  Model A, which is predominantly a four top signal,  has significantly more multi-lepton and $b$-jet events passing selection cuts than Model B and Model C, which  have fewer four top events. In Table~\ref{tab:counts}, the signal significance achievable with
$1~fb^{-1}$ integrated luminosity is shown.  By requesting at least $4$ $b$-tagged jets it  is possible to observe signal significance $S/\sqrt{B}\ge$5 for events with a single lepton.  The one-lepton four-b-jet channel will prove to be robust and the best channel for discovery.

\begin{table}[t!]

%%%%%TABLE A%%%%%

\begin{centering}
~\\
 \textbf{Number of Background Events} ($\mathbf{B}$)

\begin{tabular}{c}
 \textbf{Standard Model} \\
\begin{centering}
\begin{tabular}{cccc}
	{\bf{B}}	& \emph{2b} & \emph{3b} & \emph{4b} \\
 \emph{1$\ell$}  & 286.2 & 41.4 & 1.04  \\
 \emph{OS} & 32.8 & 5.65 & 0.007  \\
 \emph{SS}& 0.3 & 0.06 & 0 \\
 \emph{3L} & 0.14 & 0.007 & 0 \\
\end{tabular}
\end{centering}
\end{tabular}

\end{centering}

%%%%%TABLE B%%%%%

\begin{centering}
~\\
 \textbf{Number of Signal Events} ($\mathbf{S}$)

\begin{tabular}{ccc}
 \textbf{Model A} &  \textbf{Model B} & \textbf{Model C} \\

\begin{tabular}{cccc}
  {\bf{S}} & \emph{2b} & \emph{3b} & \emph{$\geq $4b} \\
  \emph{1L} & 47.1 & 39.3 & 19.3\\
  \emph{OS} & 12.4 & 9.9 & 3.9\\
  \emph{SS} & 6.6 & 5.1 & 2.3\\
  \emph{3L} & 3.0 & 2.1 & 0.7
\end{tabular}
 &
\begin{tabular}{cccc}
  & \emph{2b} & \emph{3b} & \emph{$\geq $4b} \\
  \emph{1L} & 33.5 & 26.9 & 13.8\\
  \emph{OS} & 6.4 & 5.0 & 1.7\\
  \emph{SS} & 2.3 & 1.2 & 0.2\\
  \emph{3L} & 0.7 & 1.0 & 0.3
\end{tabular}
 &
\begin{tabular}{cccc}
  & \emph{2b} & \emph{3b} & \emph{$\geq $4b} \\
 \emph{1L}  & 18.0  & 14.4 & 7.4 \\
 \emph{OS}  & 2.0  & 0.9 & 0.6 \\
 \emph{SS}  & 0.7  & 0.6 & 0.2 \\
 \emph{3L}  & 0  & 0.1 & 0.1 \\
\end{tabular}
\end{tabular}

\end{centering}

%%%%%TABLE C%%%%%

\begin{centering}
~\\
 \textbf{Significance} $\mathbf{\left(S/\sqrt{B+1}\right)}$

\begin{tabular}{ccc}
 \textbf{Model A} & \textbf{Model B}  & \textbf{Model C} \\

\begin{tabular}{cccc}
  & \emph{2b} & \emph{3b} & \emph{$\geq $4b} \\
 \emph{1L}  & 2.77  & 6.03 & 13.5 \\
 \emph{OS}  & 2.13  & 3.83 & 3.88 \\
 \emph{SS}  & 5.75  & 4.95 & 2.30 \\
 \emph{3L}  & 2.80  & 2.09 & 0.70 \\ 
\end{tabular}
 &

\begin{tabular}{cccc}
  & \emph{2b} & \emph{3b} & \emph{$\geq $4b} \\
 \emph{1L}  & 1.97  & 4.13 & 9.66 \\
 \emph{OS}  & 1.10  & 1.93 & 1.69 \\
 \emph{SS}  & 2.00  & 1.16 & 0.20 \\
 \emph{3L}  & 0.65  & 0.99 & 0.30 \\
\end{tabular}
 &
\begin{tabular}{cccc}
  & \emph{2b} & \emph{3b} & \emph{$\geq $4b} \\
 \emph{1L}  & 1.06  & 2.21 & 5.18 \\
 \emph{OS}  & 0.34  & 0.34 & 0.40 \\
 \emph{SS}  & 0.58  & 0.58 & 0.20 \\
 \emph{3L}  & 0  & 0.10 & 0.10
\end{tabular}

\end{tabular}

\end{centering}

\caption{\label{tab:counts}
Number of SM events, number of signal event, and signal significance, with 2, 3, or 4 b-tagged jets and $OS$, $SS$, or $3$ leptons at the early LHC-7, for $1 fb^{-1}$ integrated luminosity. For the 1-lepton counts, \emph{cut-1} was applied, while for the other lepton counts \emph{cut-2} was applied. These numbers were found for $m_{\tilde{g}} = 500$ GeV and $m_{LSP} = 100 $ GeV.}

\end{table}

\section{Scan and Results}

\begin{figure}[th!]

\includegraphics[scale=.3]{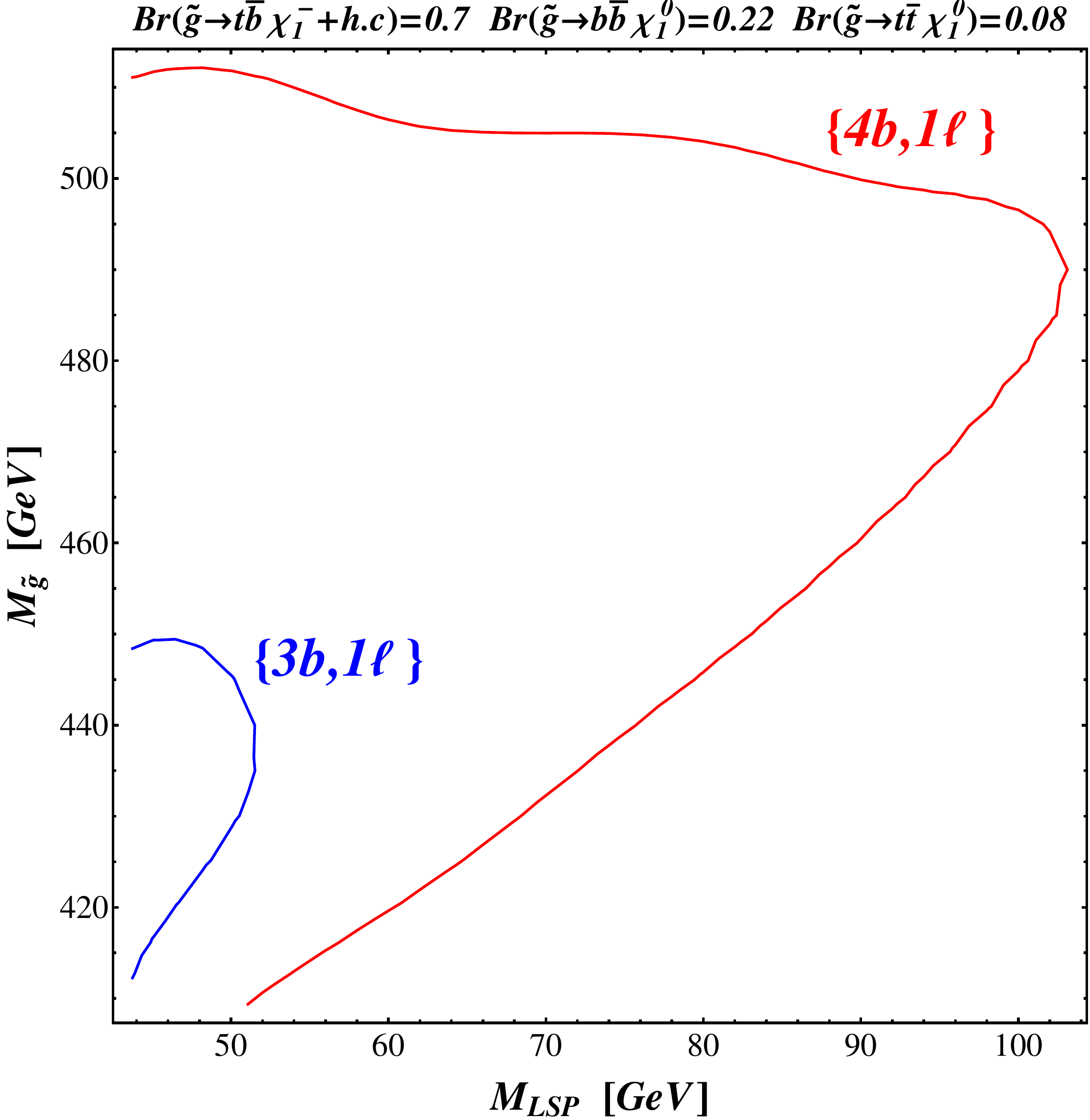} 
%\end{figure}

%\begin{figure}[th!]
\includegraphics[scale=.3]{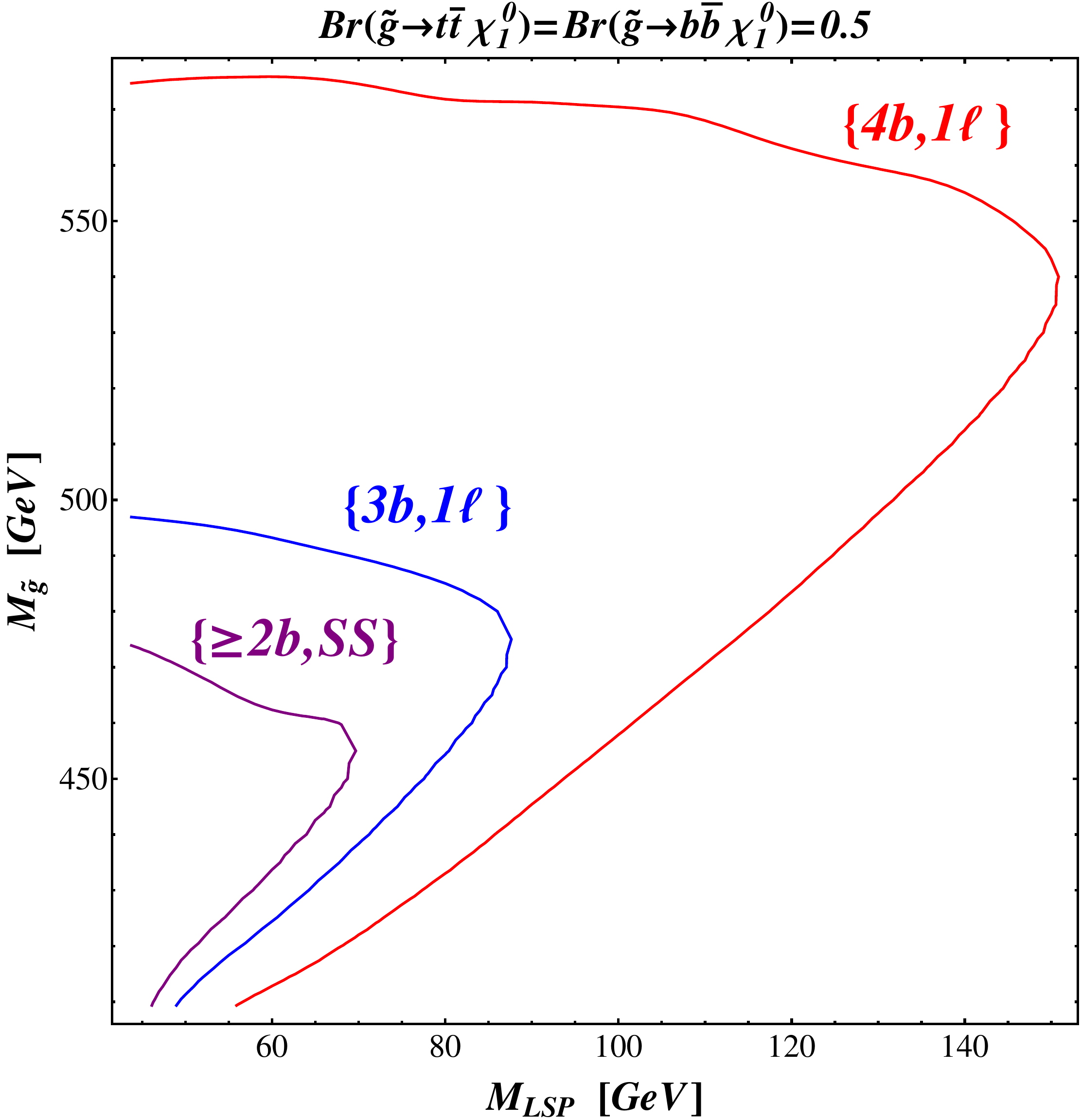} 
\caption{\label{fig:ttttmultiplechannels}$\sigma=5$  contours  $\{ b,\ell \}$-channels at LHC-7 TeV for $1 fb^{-1}$ integrated luminosity of gluino pair production.  In all Models, the $\{4 b, 1 \ell \}$- channel provides the best channel for discovery. In model A, where all events contain four tops,  the SS-dilepton channel can be a competitive mode for discovery. In all models, there are other channels that will give a lower but noticeable excess, and will provide a valuable confirmation of a mutli-top signal. \vspace{2.25in}}
\end{figure}

For each model (a fixed  $m_{\tilde{g}}$ and $m_{LSP}$), we simulated $1 fb^{-1}$ of data using Pythia and PGS. Then we searched for the models over the backgrounds for the selection cuts  in Eqs.~\ref{cut1}-\ref{cut2} in each of the $b$-jet and lepton ($\{ b,l \}$) channels. A statistical significance in a $\{b,\ell\}$ channel is defined as $\sigma_{\{ b,\ell \}   } \equiv \frac{S_{ \{ b,\ell \}   } }{\sqrt{B_{ \{ b,\ell \} }+1}}$
where $S_{\{ b,l \}   }  (B_{\{ b,\ell \}   })$ is the number of signal(background) events expected to be in the $\{ b,\ell \}$-channel for one of the two selection cuts  in Eqs.~\ref{cut1}-\ref{cut2}. Thus, if for any of the significances,   $\sigma_{cut_i,  \{ b,\ell \}   }\ge 5$, the model can be considered discoverable at $1 fb^{-1}$. In Figures \ref{fig:ttttmultiplechannels} we plot $\sigma_{cut_1,  \{ b,\ell \}   }= 5$ contours, for the channels
\[ \{ \ge 4b, 1 \ell \} ~~~~ \{ 3b, 1 \ell \} ~~~~ \{ \ge 2b, SS \} ~~~~ \{ \ge 2b, OS \} ~~~~ \{ \ge 1b, 3 \ell \}  .\]
In the first two channels \emph{cut-1} is used, and in the last three channel, the weaker \emph{cut-2}, is used. As is evident from Table \ref{tab:counts}, the backgrounds for $\{ \ge 4b, 1 \ell \}$ are significantly smaller than the backgrounds for $\{ 3b, 1 \ell \}$, and therefore it is not beneficial to combine them into the inclusive channel $\{ \ge 3b, 1 \ell \}$. The channels we used in this study maximize the significance.

In all case the $\{\ge 4b, 1 \ell \}$- channel provides the best channel for discovery. But, the SS-dilepton channel can be a competitive mode for discovery. It is important that the 4-top final state will give signatures in several channels if it appears in any.  Finding a second predicted channel would be valuable confirmation.  If two or more channels are present a combined significance would be a useful construct and facilitate a claim of discovery.

\section{Summary}
We have studied the signatures of low energy supersymmetry in multi-top and/or multi-$b$ production at 7 TeV LHC, and associated Standard Model backgrounds. Results are presented in terms of discovery reaches for $1 fb^{-1}$. In recent years a number of models have been proposed that lead to such final states. 
The required spectrum, heavy squarks with the third generation somewhat lighter than the first two and  light gluino, satisfies the existing experimental constraints better and can be motivated on very general theoretical grounds. 
In addition, it has been realized that generic string theories compactified to 4 dimensions and satisfying phenomenological constraints typically lead to such final states (as briefly described in the introduction). Thus such final states have emerged as an unusually well-motivated discovery channel at LHC. We focus on gluino pair production in supersymmetric theories both because of the strong theoretical motivations and because of the well defined nature of the such models. At 7 TeV LHC with $1 fb^{-1}$ the reach can be over 600 GeV (up to about 650 GeV) gluino mass. 
Discovery reach at higher luminosity can be scaled from our result straightforwardly.  Precise discovery reach at a different energy requires a different full study, such as the case of $E_{\rm cm}=14$ TeV studied in Ref.~\cite{14TeVPaper}. However, we can roughly estimate for $E_{\rm cm} = 8$ TeV, the reach in gluino mass can be enhanced by about a factor of $8/7$. 
Top reconstruction was studied in \cite{14TeVPaper} and is difficult, but counting leptons and $b$-jets excess for discovery is robust. The size of the counting signal provides information on the gluino cross section, which in turn is correlated with the gluino spin. 
Addition kinematical distributions could also help to enhance the discovery reach. More careful analysis, preferably with data driven approaches, will be necessary to understand the background distribution in detail. We urge experimentalists to focus attention on these channels.

\section{Acknowledgments}
We thank Bobby Acharya, Daniel Feldman, Brent Nelson and Aaron Pierce for discussion. E.K. is grateful to the String Vacuum Project for travel support and for a  String Vacuum Project Graduate Fellowship funded through NSF grant PHY/0917807. This work was supported by the DOE Grant \#DE-FG02-95ER40899. L.-T.W. is supported by the NSF under grant PHY-
0756966, and by a DOE Early Career award under grant \#DE-SC0003930.

While this work was in preparation for posting \cite{Gregoire:2011ka} appeared which explores similar issues.
%\newpage

\begin{figure}[th!]
\includegraphics[scale=.3]{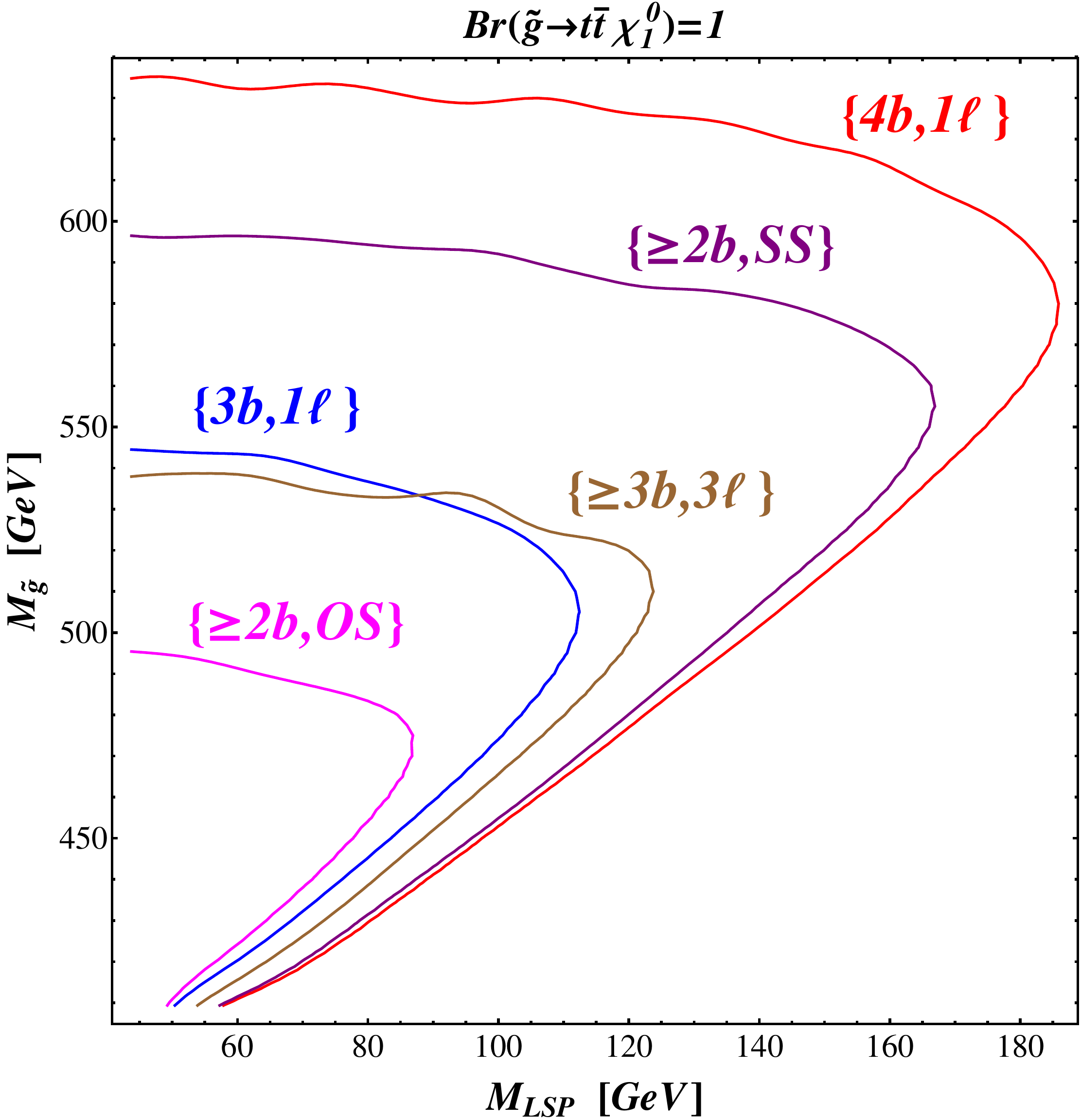} 
\end{figure}

{}

\end{document}